\documentclass[twocolumn,floatfix,preprintnumbers,nofootinbib,superscriptaddress]{revtex4}

\usepackage{ulem}
\usepackage{bm}
\usepackage{times}
\usepackage{amssymb,amsbsy,amsmath,amsfonts}
\usepackage{graphicx}
\usepackage{float}
\usepackage{color}
\usepackage{morefloats}
\usepackage{rotating}
\usepackage{srcltx}
\usepackage{slashed}
\usepackage{subfigure}
\usepackage{multirow}
\usepackage{verbatim}
\usepackage{hyperref}
\usepackage{tabularx}
\usepackage{adjustbox}
\usepackage{array}
\usepackage{tabularray}

\newcommand{\HH}{${}^4_\Lambda$H\@}

\newcommand{\HHH}{${}^3_\Lambda$H\@}

\begin{document}
\title{Perspectives for hyperon and hypernuclei physics}
\author{Jin-Hui Chen} \email{chenjinhui@fudan.edu.cn}
\affiliation{Key Laboratory of Nuclear Physics and Ion-beam Application (MOE), and Institute of Modern Physics, Fudan University, Shanghai 200433, China}

\author{Li-Sheng Geng} \email{lisheng.geng@buaa.edu.cn}
\affiliation{School of Physics, Beihang University, Beijing 102206, China}
\affiliation{Peng Huanwu Collaborative Center for Research and Education, Beihang University, Beijing 100191, China}
\affiliation{Sino-French Carbon Neutrality Research Center, \'Ecole Centrale de P\'ekin/School of General Engineering, Beihang University, Beijing 100191, China}
\affiliation{Beijing Key Laboratory of Advanced Nuclear Materials and Physics, Beihang University, Beijing 102206, China }
\affiliation{Southern Center for Nuclear-Science Theory (SCNT), Institute of Modern Physics, Chinese Academy of Sciences, Huizhou 516000, China}

\author{Emiko Hiyama} \email{hiyama@riken.jp}
\affiliation{Department of Physics, Graduate School of Science, Tohoku University, Sendai 980-8578, Japan}
\affiliation{RIKEN Nishina Center for Accelerator-Based Science, Wako, Saitama 351-0198, Japan}

\author{Zhi-Wei Liu} \email{liuzhw@buaa.edu.cn}
\affiliation{School of Physics, Beihang University, Beijing 102206, China}

\author{Josef Pochodzalla} \email{pochodza@uni-mainz.de}
\affiliation{Institut für Kernphysik, Johannes Gutenburg-Universität, D-55099, Mainz, Germany}
\affiliation{Helmholtz-Institut Mainz, Johannes Gutenberg-Universit\"at Mainz, 55099 Mainz, Germany}
\affiliation{PRISMA$^+$ Cluster of Excellence, Johannes Gutenberg-Universit\"at Mainz, 55099 Mainz, Germany}
\begin{abstract}
Hypernuclei, nuclei containing one or more hyperons, serve as unique laboratories for probing the non-perturbative quantum chromodynamics (QCD). Recent progress in hypernuclear physics, driven by advanced experimental techniques and theoretical innovations, is briefly reviewed with a focus on key findings and unresolved challenges, such as the precise determination of the hypertriton binding energy, investigations of charge symmetry breaking in mirror hypernuclei, and the search for exotic systems, including the neutral nn$\Lambda$ state. Experimental breakthroughs, including invariant-mass analyses and femtoscopy studies in heavy-ion collisions, as well as high-resolution $\gamma$-spectroscopy, have enabled precise studies of light hypernuclei and offered critical insights into the hyperon-nucleon interaction. Theoretical progress, including ab initio calculations based on chiral effective field theory and lattice QCD, has further enhanced our understanding of hyperon-nucleon and hyperon-hyperon interactions.
\end{abstract}

\pacs{}
\date{\today}

\maketitle

\date{\today}
\maketitle

\section{Introduction}\label{sec1}
Nuclei containing strange baryons, so-called Hypernuclei, are unique femto-laboratories for multi-baryon interactions with hyperons~\cite{Davis:2005mb, Dalitz:2005mc,Gal:2016boi, Chen:2018tnh, Tolos:2020aln}. Light hypernuclei are particularly interesting, as not only phenomenological models but also ab initio studies based on chiral effective field theories and even lattice quantum chromodynamics calculations are available for such systems~\cite{PhysRevC.93.014001, NPLQCD:2012mex, Petschauer:2020urh}.

During the first two decades of hypernuclear research, nuclear emulsions were the main source of information about hypernuclei. Even today, data from emulsions still provide the most precise information on the binding energy of many hypernuclei. Around the turn of the century, high-resolution $\gamma$-spectroscopy of hypernuclei with germanium detectors became the most important tool for decay studies. These measurements provide precise information on the level schemes of various hypernuclei, enabling the extraction of different spin-dependent components of the $\Lambda$-nucleon interaction. On the other hand, employing quasi-two-body kinematics, ground and excited hypernuclear states can be identified through a missing-mass analysis of the incident beam and the observed associated meson. Since these reactions require stable target nuclei, the hypernuclei accessible by these reactions lie on or near the $\beta$-stability line. In the future, this limitation can be overcome by heavy-ion reactions, where particularly projectile fragmentation with secondary beams can provide access to proton or neutron-rich hypernuclei. In addition to strange nuclei, exotic atoms with strange hadrons also provide a wealth of information for non-perturbative QCD studies at threshold~\cite{Dalitz:1960du, ExHIC:2017smd, Guo:2017jvc, Chen:2024eaq}.

\begin{figure}[!b]
\begin{center}
\includegraphics[width=0.50\textwidth]{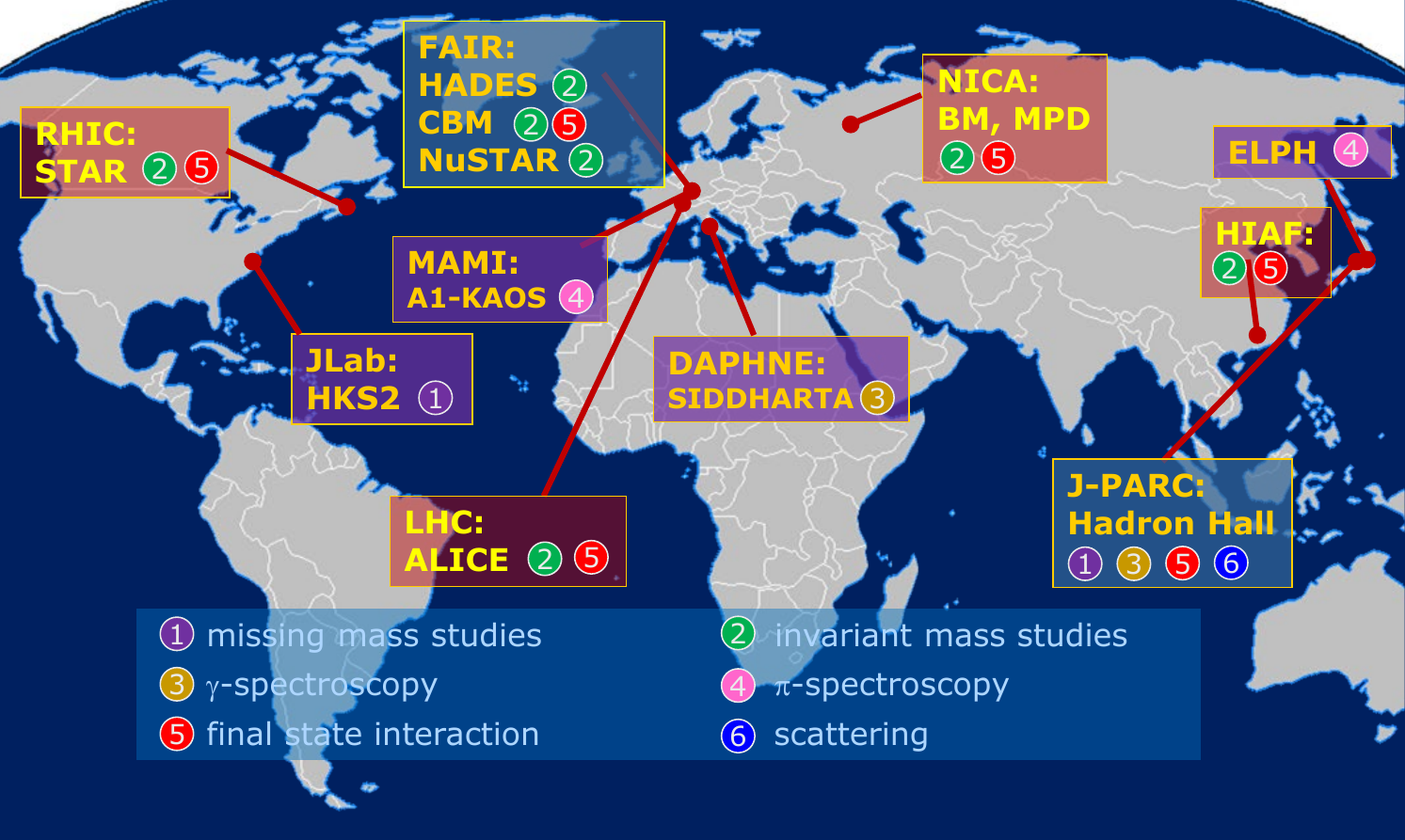}
\end{center}
  \caption{Global planned or operational facilities studying properties of strange nuclear systems.
  }
\label{fig:worldmap}
\end{figure}

Although several new experimental techniques have been developed in hypernuclear physics over the past decade, our understanding remains limited to a few hypernuclei.
Furthermore, the number of observed events is often rather low, and in many cases, the resolution is also limited.
Fortunately, several experiments are planned or already underway at various laboratories worldwide (Fig.~\ref{fig:worldmap}) that will help usher in the era of high precision and high statistics in strangeness nuclear physics. The complementarity of these different experimental approaches, combined with guidance from state-of-the-art theories, will provide a solid basis for a comprehensive understanding of strange hadrons in cold hadronic matter.

\section{Future experimental highlights}
\subsection{The lightest hypernucleus {\HHH}}

The lightest strange baryonic nuclear system is the hypertriton \HHH\ and
found to be just bound.
For \HHH\, accessible observables are in principle the binding energy, the excitation spectrum (if particle-bound excited states exist), spins, lifetime, and decay branching ratios. Of course, for such a light nucleus, these quantities are expected to be intimately related~\cite{DALITZ196258,Rayet1966}.

Like the deuterium or the tritium nucleus for conventional baryon interactions, a profound understanding of the lightest hypernuclei is a cornerstone of any strong interaction theory dealing with strange baryons~\cite{Downs:1959zz,STAR:2010gyg,Chen:2023mel}. For fitting modern interactions, the \HHH\ is used to determine the spin dependence~\cite{Kamada:1997rv,Sun:2025oib}.
A precise and accurate measurement of the hypertriton binding energy gains further relevance from recent hypernuclear studies in ultra-relativistic heavy-ion collisions~\cite{STARmass2020,PhysRevLett.131.102302}. The large size of the 'ultimate halo nucleus' {\HHH} makes it a unique probe for the fragment formation process in ultra-relativistic heavy-ion collisions.

Less than 10 years ago, hypertriton emulsion data were the only source of its binding energy $B_{\Lambda}$.
The number of uniquely identified events for the determination of $B_{\Lambda}$ was of the order $\sim$200 from different decay modes, analyzed and compiled in three works. The mean
value was evaluated in Ref.~\cite{Juric:1973zq} using both decay modes (0.13$\pm$0.05)\,MeV. Recently, two new values became available, one by the STAR Collaboration~\cite{STARmass2020},
$B_{\Lambda} = 406 \pm 120\, \mathrm{(stat.)} \pm 110\, \mathrm{(syst.)}$\,keV,
and the other by the ALICE Collaboration~\cite{PhysRevLett.131.102302},
$B_{\Lambda} = 102 \pm 63\, \mathrm{(stat.)} \pm 67\, \mathrm{(syst.)}$\,keV.
Both were determined from relativistic heavy-ion experiments with large sets of collected collisions.
Remarkably, the STAR B$_{\Lambda}$ value is 
larger than the ALICE one, and they differ by two standard deviations. The STAR value also seems to be in tension with the emulsion value. Ref.~\cite{Chen:2023mel} presents a global average of all data points, yielding $B_{\Lambda} = 0.17 \pm 0.06$\,MeV. Note that $S\equiv\sqrt{\chi^2/(N-1)}=1.6$ from the average is rather large~\cite{Chen:2023mel}, indicating a relatively large spread of experimental data. The ideogram presentation of the data points in Ref.~\cite{Chen:2023mel} also shows a large spread of the values. A thorough understanding of the systematics is required~\cite {Chen:2023mel}.

Several experiments are expected to produce more precise and accurate hypertriton lifetime and binding energy data in the coming years. The WASA-FRS Collaboration at GSI/FAIR, the P73/P77 Collaboration at J-PARC, ALICE during LHC Runs 3 and 4, and an experiment at ELPH at Tohoku University will enhance our understanding of the lifetimes of {\HHH} and {\HH}.
One activity of the Extension Project for the J-PARC Hadron Experimental Facility
is focusing on the $\gamma$-ray spectroscopy of hypernuclei and will search for the $\gamma$-ray transitions of {\HHH}. 
A unique experiment at the R3B spectrometer of GSI/FAIR will probe the $\Lambda$ halo structure of {\HHH}.
New measurements of the {\HHH} binding energy are expected from the Mainz Mikrotron (MAMI), JLab, ALICE, and the J-PARC E07 emulsion experiment. These attempts attest to the importance of the hypertriton as a benchmark for any hypernuclear structure calculation.

\subsection{Hypernuclei produced from radioactive beams and new methods}
Heavy-ion collisions at GeV/nucleon energies are a powerful tool for producing and studying hypernuclei.
GSI/FAIR is a unique facility in Europe for performing hypernuclear studies, as it provides high beam rates at energies above the strangeness production threshold for both stable and unstable beams.
The (Super-)FRS and R3B collaborations of NUSTAR offer appealing capabilities for investigating the structure of light hypernuclei using the invariant-mass technique.
Following the pioneering HypHI experiment, the FRS collaboration initiated a program combining the WASA detector with the FRS spectrometer. The R3B collaboration proposes a complementary investigation based on a time-projection tracker inside the GLAD magnet. Neutron detection in coincidence is made possible by the NeuLAND array.
Both programs should facilitate further investigation of the properties of hypernuclei, including neutron-deficient hypernuclei, which can be produced more efficiently using high-intensity radioactive beams.

Neutron-rich hypernuclei represent a new avenue to be explored in the coming decades. The re-acceleration to 4-5 GeV/nucleon of high-intensity radioactive beams from the Super-FRS is an exciting long-term perspective to investigate for the future of FAIR. Furthermore, neutron-rich hypernuclei could be reached by charge exchange reactions, where one or two charge exchange processes follow the production of a hypernucleus.

The production of strangeness in heavy-ion collisions at these energies can be further investigated at the HADES experiment at GSI/FAIR. Light-ion collisions at energies close to the production threshold (1.6 GeV/nucleon) and higher energies would allow to measure differential production cross section for $\Lambda$ hyperons, {\HHH} and {\HH}, benchmarking cascade models towards a refined understanding of the production mechanism, a key ingredient to prepare dedicated hypernuclei experiments at GSI/FAIR.

Low-energy antiprotons forming antiprotonic atoms open a new way to produce hypernuclei and perform their spectroscopy with high energy resolution. About 3\% of antiproton-nucleon annihilations occurring at the nuclear surface 
produce strangeness, mostly kaons. These relatively low-energy kaons are likely to re-interact with the residual nucleus and produce a $\Lambda$ hyperon. Theoretical estimates suggest that approximately 1\% of annihilations lead to the production of a hypernucleus, a potential breakthrough for exploring the $\Lambda$ hypernuclear landscape at the ELENA facility within the Antiproton Decelerator at CERN.

\subsection{The case of the neutral nn$\Lambda$ system}
\label{sec:nnl}

The small binding energy of the hypertriton leads to predictions of the non-existence of bound hypernuclei for isotriplet three-body systems such as nn$\Lambda$. However, invariant mass spectroscopy at GSI has reported events that may be interpreted as the bound nn$\Lambda$ state~\cite{PhysRevC.88.041001}. Consistent with the observed lifetime of this system, typical for a weak decay, the measured binding energy is about 1\,MeV.
Theoretically, this neutral hypernucleus is likely to be unbound, though no firm predictions of its stability can be made. An experimental clarification is required.

At J-Lab, the nn$\Lambda$ state was sought by missing-mass spectroscopy via the $^3$H(e,e'K$^+$) reaction at Jefferson Lab's experimental Hall A \cite{10.1093/ptep/ptab158}. No significant structures were observed close to B$_{\Lambda}$=0\,MeV.  Therefore, only upper limits of the production cross-section of a nn$\Lambda$ state were obtained. These limits depend moderately on the position and width of a possible nn$\Lambda$ state~\cite{10.1093/ptep/ptab158}.

Following tentative observation at GSI~\cite{HypHI:2013sxa}, the search for the nn$\Lambda$ system in the $^6$Li+$^{12}$C reaction was repeated with an improved setup at GSI, utilizing the fragment separator facility and the WASA detector. The experiment was successfully performed in 2022.
The WASA detector array was originally used at the Svedberg laboratory in Sweden and later at the COSY ring at Forschungszentrum Jülich. Over the past few years, the WASA detector has been installed at the Fragment Separator (FRS) at GSI/FAIR. It was equipped with a new mini drift chamber and additional beam counters. The WASA detector can be opened in the middle to make space for the superconducting solenoid magnet.
The FRS/WASA is complementary to the time-projection tracker inside the GLAD magnet.

The goal of the experiment is to study light hypernuclei during FAIR Phase 0. The experiment is looking for weak mesonic decays of light hypernuclei produced in $^6$Li+$^{12}$C and $^{12}$C+$^{12}$C reactions at a beam energy of 1.96 GeV/nucleon. The analysis is carried out using the invariant mass method~\cite{HypHI:2013sxa}. The $^{12}$C target is positioned in the mid-focal plane of the FRS. Produced hypernuclei decay weakly behind the primary target. Charged pions are detected on the upgraded WASA detector. The residual nucleus (in the case of the hypertriton $^{3}$He) is emitted in the forward direction and is uniquely identified in the second part of the FRS. Thus, the setting of the FRS separator determines the selected hypernucleus.

With a $^6$Li beam at an energy of 1.96 GeV/nucleon, three different settings of the fragment separator can be measured, selecting $^3$He, $D$, or $^4$He. These settings correspond to the weak pionic decays of {\HHH}, nn$\Lambda$ or {\HH}~\cite{HypHI:2013sxa}. In addition, with a $^{12}$C beam, one more setting can be studied, selecting $^3$He and simultaneously $^9$C, corresponding to the decay of {\HHH} or {$_{\Lambda}^{9}$B}. The analysis has started, and the first results are expected soon. 

\subsection{Study of charge symmetry breaking in hypernuclei}
The charge symmetry breaking (CSB) was quite small in the $NN$ interaction~\cite{Miller:1990iz}. One of the most famous CSB effects was seen in the $A=3$ iso-doublet nuclei $^3$H and $^3$He. A 70-keV difference between their binding energies was found after removing the contributions from Coulomb and magnetic effects, which is regarded as the breaking of charge symmetry. This effect can be explained by the mixing effects of $\pi^0\eta$ and $\rho^0\omega$ in the $NN$ interaction.

The charge symmetry should also appear in the interaction between the strange baryon $\Lambda$ and a nucleus, since $\Lambda$ is charge-free. Unfortunately, the scattering lengths of $\Lambda p$ and $\Lambda n$ cannot be easily determined due to the difficulties of scattering experiments. However, it can also be studied by measuring the $\Lambda$ binding energies of hypernuclei. By adding one $\Lambda$ to $^3$H and $^3$He, respectively, a pair of $A=4$ mirror hypernuclei, $^4_{\Lambda}$H and $^4_{\Lambda}$He, are formed. The difference in their $\Lambda$ binding energies of ground states ($0^+$) was measured to be $\Delta B^{4}_{\Lambda}(0^+)=B_{\Lambda}(^{4}_{\Lambda}{\rm He})-B_{\Lambda}(^{4}_{\Lambda}{\rm H})=350\pm40$~keV in nuclear emulsion experiments from 1970s~\cite{Juric:1973zq}. This CSB effect is significantly larger than that in $^3$H-$^3$He, indicating that the $\Lambda p$ interaction is stronger than the $\Lambda n$ interaction. In 2015, the J-PARC E13 collaboration measured the energy of $\gamma$ ray emitted from excited $^4_{\Lambda}$He precisely~\cite{J-PARCE13:2015uwb}. They found that the CSB effect in $^4_{\Lambda}$H and $^4_{\Lambda}$He excited states ($1^+$) is close to zero $\Delta B^{4}_{\Lambda}(1^+)=-30\pm50$~keV, which is much smaller than that in their ground states. These experimental results indicate that a special mechanism in the $\Lambda$N interaction may exist.

The CSB effect is also seen in the $A=7, T=1$ hypernuclei: $^7_{\Lambda}$He($\alpha$+n+n+$\Lambda$), $^7_{\Lambda}$Li$^*$($\alpha$+p+n+$\Lambda$), and $^7_{\Lambda}$Be($\alpha$+p+p+$\Lambda$). Due to their cluster structures of $\alpha$+NN+$\Lambda$, it is possible to extract $\Lambda p$ and $\Lambda n$ binding energies from their $\Lambda$ binding energy difference: $\Delta B^{7}_{\Lambda{\rm n}}=B_{\Lambda}(^{7}_{\Lambda}{\rm He})-B_{\Lambda}(^{7}_{\Lambda}{\rm Li^*})=420\pm280$~keV and $\Delta B^{7}_{\Lambda{\rm p}}=B_{\Lambda}(^{7}_{\Lambda}{\rm Li^*})-B_{\Lambda}(^{7}_{\Lambda}{\rm Be})=100\pm90$~keV~\cite{HKSJLabE01-011:2012sgn}. This result indicates that $\Lambda$ is more strongly bound to a neutron than to a proton, which is contrary to the CSB effect in $^4_{\Lambda}$H and $^4_{\Lambda}$He, making the puzzle more complex. There is also a chance to study the CSB in the $A=6$ hypernuclei: $^6_{\Lambda}$H and $^6_{\Lambda}$He. However, the existence of $^6_{\Lambda}$H is still not verified, and the study of $^6_{\Lambda}$He is limited~\cite{Botta:2016kqd}.

On the theoretical side, the CSB puzzle also exists~\cite{Hiyama:2009ki,Gal:2015bfa,Haidenbauer:2021wld,Le:2022ikc,Schafer:2022une}. The calculations of the $\Lambda$N interaction with $\Lambda\Sigma^{0}$ mixing can not explain the $\Lambda$ binding energy difference between $^4_{\Lambda}$H and $^4_{\Lambda}$He in their ground and excited states at the same time~\cite{Gazda:2016qva,Schafer:2022une}. Conversely, this CSB effect can also constrain the scattering lengths of $\Lambda p$ and $\Lambda n$ and their interaction potentials~\cite{Haidenbauer:2021wld}. Within the $\alpha + \Lambda + N + N$ four-body cluster model, the structure of the $T=1$ iso-triplet hypernuclei, $^7_{\Lambda}$He, $^{7}_{\Lambda}$Li, and $^7_{\Lambda}$Be has been studied~\cite{Hiyama:2009ki}. It was pointed out that a difference of $-150$ keV between $^7_{\Lambda}$He and $^7_{\Lambda}$Be is expected to originate from the Coulomb force, suggesting that the difference in the $A=7$ hypernuclei system could be small~\cite{Hiyama:2009ki}.
Ab-initio calculations within the no-core shell model predicted that the CSB value for the $A = 7$ system is small and agrees with the splittings deduced from the empirical binding energies within the experimental uncertainty~\cite{Le:2022ikc}. The computed CSB for the $A = 8$ doublet is somewhat larger than the experimental value available~\cite{Le:2022ikc}.
In conclusion, precise measurements of the $\Lambda$ binding energies in light hypernuclei, particularly $^4_{\Lambda}$H and $^4_{\Lambda}$He, are necessary to understand the CSB mechanism in the $\Lambda$N interaction.

In 2015, the A1 Collaboration at Mainz Microtron measured the $\Lambda$ binding energy of ground-state $^4_{\Lambda}$H precisely with the electron beam and decay pion spectroscopy method~\cite{A1:2015isi}. The CSB effects in the $A=4$ hypernuclei system were updated to $\Delta B^{4}_{\Lambda}(0^+)=233\pm92$~keV and $\Delta B^{4}_{\Lambda}(1^+)=-83\pm93$~keV, still showing large difference between ground and excited states. The A1 collaboration carried out another experiment in 2022 to measure the $\Lambda$ binding energies of $^3_{\Lambda}$H and $^4_{\Lambda}$H with upgraded techniques to suppress uncertainties~\cite{A1Hypernuclear:2024bjx}. Their data analysis is still ongoing. The Jefferson Lab also plans to run the electron beam experiment with kaon/pion spectroscopy to measure the $\Lambda$ binding energies of hypernuclei in 2026~\cite{Achenbach:2024lpw}. Experiments to measure the $\gamma$ ray emitted from excited hypernuclei are under construction in J-PARC with kaon beams~\cite{Ukai:2022sey}. 

Hypernuclei can be produced during the hadronization process in heavy-ion collisions. It has been predicted that the yield of light hypernuclei such as $^3_{\Lambda}$H and $^4_{\Lambda}$H can reach its maximum in Au+Au collisions at $\sqrt{s_{\rm NN}}$ about 3~GeV~\cite{Steinheimer:2012tb,Chen:2023mel}. From 2018, the STAR experiment began running in fixed-target mode, extending the Au+Au collision energy to $\sqrt{s_{\rm NN}}$=3~GeV. The large production yields of light nuclei, $^3_{\Lambda}$H and $^4_{\Lambda}$H in Au+Au collisions at $\sqrt{s_{\rm NN}}$=3~GeV have been measured~\cite{STAR:2021orx,STAR:2023uxk,Chen:2024aom}. It enables the measurement of $\Lambda$ binding energies in large samples of hypernuclei. In 2022, the STAR collaboration published the measurement of CSB in the $A=4$ hypernuclei system in Au+Au collisions at $\sqrt{s_{\rm NN}}$ = 3~GeV~\cite{STAR:2022zrf}. By reconstructing $^4_{\Lambda}$H via its two-body decay channel $^4_{\Lambda}{\rm H}\to {\rm ^4He}+\pi^-$ and $^4_{\Lambda}$He via its three-body decay channel $^4_{\Lambda}{\rm He}\to {\rm ^3He}+{\rm p}+\pi^-$ in about 300 million events, their invariant mass distributions were obtained. Then their $\Lambda$ binding energy difference in ground states was extracted from their masses: $\Delta B^{4}_{\Lambda}(0^+)=160\pm140(stat.)\pm100(syst.)$~keV. Combined with the $\gamma$ ray energies from previous experiments, their $\Lambda$ binding energy difference in excited states was measured to be $\Delta B^{4}_{\Lambda}(1^+)=-160\pm140(stat.)\pm100(syst.)$~keV. This result shows that the CSB effects of $A=4$ hypernuclei have similar magnitudes but opposite signs between the ground and excited states. It is consistent with the calculation using chiral effective field theory $YN$ potentials plus a CSB effect~\cite{Gazda:2015qyt}. However, due to the low reconstruction efficiency of the three-body decay, the statistics of $^4_{\Lambda}$He are limited, resulting in large statistical uncertainties. According to this result, it remains challenging to solve the CSB puzzle. Luckily, STAR has recorded approximately 2 million events in Au+Au collisions at $\sqrt{s_{\rm NN}}$ = 3 GeV with upgraded detectors in 2021. The uncertainties of the CSB measurements can be suppressed in future studies. There are also opportunities to study neutron-rich hypernuclei using such large statistics, for example, $^6_{\Lambda}$H via its two-body decay channel $^6_{\Lambda}{\rm H}\to {\rm ^6He}+\pi^-$, and $^6_{\Lambda}$He via its two-body and three-body decay channels $^6_{\Lambda}{\rm He}\to {\rm ^6Li}+\pi^-$ and $^6_{\Lambda}{\rm He}\to {\rm ^4He}+{\rm d}+\pi^-$, in which studies of the CSB effect in neutron-rich environments can be carried out.

\subsection{Two- and multi-particle hyperon correlations}

Femtoscopy has recently developed into a flourishing field for studying two- and three-body interactions of unbound hadronic systems~\cite{Liu:2024uxn}.
Furthermore, this method also provides constraints on the temporal evolution of the emission of various particles (see, e.g., Ref.~\cite{EurPhsJour53.2025}).

Correlations at small relative momentum measured in p-p and p-Pb collisions at the most violent energy available to this end at the LHC have been exploited to validate for the first time lattice QCD calculations of the nucleon-$\phi$~\cite{ALICE:2021cpv}, nucleon-$\Xi$~\cite{LHC}, and nucleon-$\Omega$ interactions~\cite{ALICE:2020mfd}.  Very precise data for the p-$\Lambda$ and p-K correlations have complemented low statistics scattering data with highly precise measurements~\cite{ALICE:2021njx,ALICE:2019gcn}, and these measurements can be used to constrain chiral effective field theory models more precisely.

The future perspective in this field is to exploit the high statistics expected from the Run 3 campaign at the LHC and the newly upgraded high-rate ALICE detector to investigate the so far poorly understood nucleon-$\Sigma$ interaction.
The program also includes the investigation of the three-particle correlation function using the measurement of the momentum correlation for triplets such as p-p-$\Lambda$ or p-p-$\Sigma$, but also exploiting
deuteron-$\Lambda$ and deuteron-$\Sigma$ correlations to investigate the isospin dependence of the three-particle correlation function.
This method will thus complement the study of hypernuclei and provide further constraints on theoretical models and input for the equation of state (EoS) of neutron stars, where three-body forces are expected to play a fundamental role.

\subsection{Theoretical studies of hyperon-nucleon and hyperon-hyperon
interactions}
Over the years, many theoretical approaches have been developed to construct hyperon-nucleon (YN) and hyperon-hyperon (YY) interactions, including various phenomenological models, lattice Quantum Chromodynamics (QCD) simulations, and chiral effective field theories, which are based on the most general Lagrangian allowed by the assumed symmetries, particularly the (broken) chiral symmetry of QCD. This section provides a brief review of the development and latest achievements of these three methods.

Similar to the nucleon-nucleon (NN) interaction, the theoretical study of YN and YY interactions began with meson-exchange models, among which the Nijmegen meson-exchange model and the Bonn-J\"ulich meson-exchange model are the most representative. 1) Nijmegen meson-exchange model. In the 1970s, the Nijmegen collaboration constructed a series of hard-core potentials based on one-boson exchanges (including pseudoscalar mesons, vector mesons, and scalar mesons, etc.)~\cite{Nagels:1976xq}, achieving simultaneous descriptions of NN and YN scattering. In 1978, Nagels et al. utilized the Regge-pole theory to construct the nucleon-nucleon potential, obtaining nuclear forces based on one-boson exchanges~\cite{Nagels:1977ze}. In 1989, Maessen et al. extended this to the hyperon-nucleon system using SU(3) flavor symmetry~\cite{Maessen:1989sx}. Due to its softer behavior near the origin, it was termed the soft-core model (NSC89). In 1997, Rijken et al. considered SU(3) flavor symmetry-breaking effects, introduced mass cutoffs at interaction vertices, and treated the coupling constant ratio $F/(F+D)$ as an adjustable parameter, refining and developing the soft-core model and proposing the NSC97 (a-f) models~\cite{Rijken:1998yy}. These models can accurately describe NN and YN scattering data simultaneously and enable reliable calculations for the well depths of 28 MeV in $\Lambda$ hypernuclei~\cite{Rijken:1998yy,Millener:1988hp,Hasegawa:1996fj}. Since then, the Nijmegen collaboration has continuously updated and developed the soft-core model~\cite{Nagels:2023zky}. 2) Bonn-J\"ulich meson-exchange model. In 1987, the Bonn collaboration proposed the famous Bonn potential model, achieving a high-precision description of NN scattering data. In 1989, Holzenkamp et al. extended the Bonn potential to the YN system, constructing the J\"ulich89 model~\cite{Holzenkamp:1989tq}. The subsequently updated J\"ulich94 version removed the energy-dependent terms in the previous potential~\cite{Reuber:1993ip}, facilitating convenient and reliable application in the first-order Brueckner calculations of $\Lambda$ single-particle potentials. In 2005, based on this, Haidenbauer and Mei\ss{}ner constrained the $\sigma$ and $\rho$ exchange potentials using a microscopic model of $\pi\pi$ and $K\bar{K}$ exchanges and phenomenologically considered short-range contributions from $a_0$ and $\kappa$, proposing the J\"ulich04 model~\cite{Haidenbauer:2005zh}, which has been also successfully applied in scattering experiments and studies of hypernuclear structures.

Attempts have also been made to describe hyperon-nucleon interactions using quark models. In 1988, the Beijing-T\"ubingen collaboration proposed a quark cluster model based on one-gluon, pseudoscalar meson, and phenomenological $\sigma$ meson exchanges, utilizing the resonating group method to treat 6-quark scattering processes~\cite{Straub:1988gj}. Based on this method, the modified quark model proposed by Fernández et al. was successfully applied in nuclear force studies~\cite{Fernandez:1993hx}. Subsequently, the Beijing-T\"ubingen collaboration extended the modified quark model to the YN system~\cite{Zhang:1994pp}, simultaneously achieving good descriptions of the NN and YN systems. In 1992, Wang and colleagues proposed the quark delocalization and color screening model by considering fully and incompletely confined configurations in Hilbert space and distinguishing between quark interactions inside and outside hadrons~\cite{Wang:1992wi}. This model addressed the lack of intermediate attractive interactions in previous quark models, successfully describing dibaryon interactions and exhibiting certain predictive capabilities. Another quark model is the Kyoto-Niigata SU(6) quark model proposed by Fujiwara et al. in 1995~\cite{Fujiwara:1995td}. It is based on the full Fermi-Breit interaction, introduces flavor symmetry-breaking effects, and utilizes the resonating group method to treat short-range interactions. Mid- and long-range interactions are modeled using the Nijmegen meson-exchange scheme. Subsequently, the SU(6) quark model has undergone continuous development~\cite {Fujiwara:2006yh}.

Following the 1990s, lattice QCD simulations gained popularity with the increasing availability of computing power and the development of more advanced numerical algorithms. While significant progress has been made in recent years, high-precision calculations of baryon-baryon interactions at physical pion mass and in large volumes remain computationally challenging. They are an ongoing pursuit in the field. The NPLQCD collaboration and the HAL QCD collaboration are the two main groups studying baryon-baryon interactions using lattice QCD simulations. In 2005, the NPLQCD collaboration proposed using the L\"uscher finite-volume method to study YN interactions~\cite{Beane:2003yx}. In 2009, the HAL QCD collaboration employed the HAL QCD method, where nonlocal potentials are defined within the Nambu-Bethe-Salpeter wave function, to calculate the $S$-wave potential and scattering length for proton-$\Xi^0$ hyperons at $m_\pi = 370, 510$ MeV~\cite{Nemura:2008sp}. Since then, these two collaborations have conducted extensive lattice QCD simulations on dibaryon systems in the strangeness sector, but all of these simulations have used a pion mass ($m_\pi$) larger than 300 MeV. Recently, the HAL QCD collaboration has achieved lattice simulations for dibaryon systems near the physical pion mass ($m_\pi = 146$ MeV), such as the $\Lambda\Lambda-\Xi N$ system~\cite{HALQCD:2019wsz}.

As a significant milestone in the development of nuclear forces, Weinberg proposed constructing the nucleon-nucleon (NN) interaction using chiral perturbation theory, known as the chiral nuclear force, which has achieved tremendous success. As a natural extension of the NN interaction to the SU(3) flavor space, constructing YN and YY interactions based on chiral effective field theories also holds great potential. In 2005, inspired by the successful study of nuclear forces using the heavy-baryon (HB) chiral effective field theory (ChEFT), the Bonn-J\"ulich collaboration employed this method to calculate YN and YY scattering observables at leading order (LO)~\cite{Polinder:2006zh, Polinder:2007mp}. The LO potential consists of four-baryon contact terms without derivatives and pseudoscalar meson exchange terms, containing five low-energy constants to be determined. Based on this chiral potential, the scattering amplitude can be obtained by solving the scattering equation nonperturbatively, where renormalization is achieved by introducing an exponential regulator function. The results reproduce the data but with slightly lower precision compared to the phenomenological models such as NSC97f and J\"ulich04~\cite{Rijken:1998yy,Haidenbauer:2005zh}. By further requiring the contact-term potential to satisfy strict SU(3) flavor symmetry, the Bonn-J\"ulich collaboration constructed leading-order chiral potentials for systems with $S = -2$, $-3$, $-4$, $-5$, and $-6$~\cite{Polinder:2007mp, Haidenbauer:2009qn, Haidenbauer:2017sws}. In 2013, Haidenbauer et al. further constructed YN interactions at the next-to-leading order (NLO)~\cite{Haidenbauer:2013oca, Haidenbauer:2015zqb}, achieving a precision comparable to that of established phenomenological models, such as NSC97f and Jülich04. In 2023, Haidenbauer et al. further advanced the YN chiral interaction to the next-to-next-to-leading order (N$^2$LO) by adopting a novel regularization scheme and considering the SU(3) flavor symmetry breaking, which achieved a good description of YN scattering data~\cite{Haidenbauer:2023qhf}. Recently, a study of light $\Lambda$ hypernuclei was carried out based on the N$^2$LO YN interactions and the no-core shell model~\cite{Le:2024rkd}, where the chiral $\Lambda NN$ and $\Sigma NN$ three-body forces were treated consistently for the first time.

\begin{figure}
\begin{center}
\includegraphics[scale=0.35]{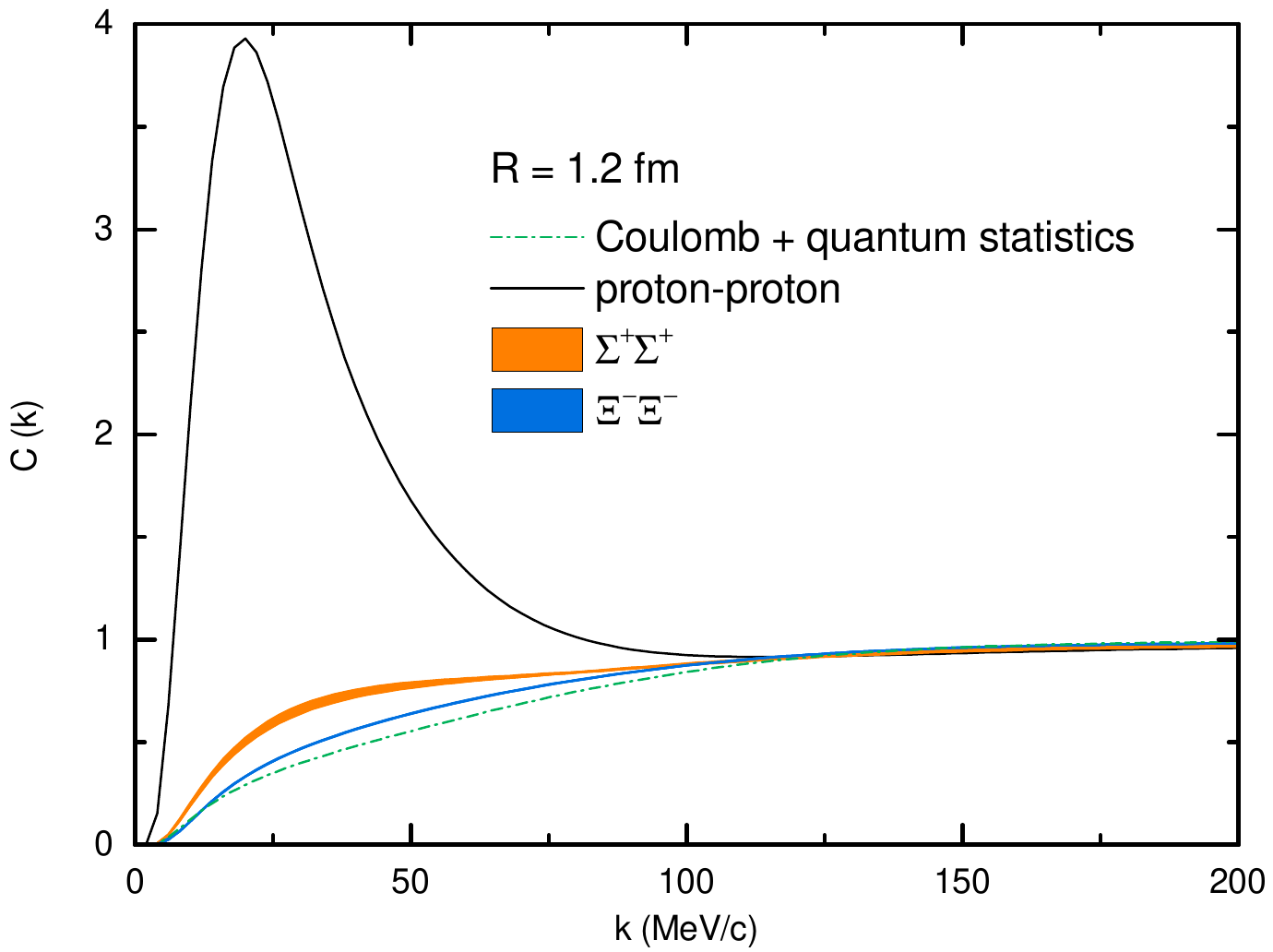}
\caption{Predicted proton-proton, $\Sigma^+\Sigma^+$, and $\Xi^-\Xi^-$ femtoscopic correlation functions as a function of the relative momentum $k$. The results are calculated with the LO relativistic baryon-baryon interactions and with a Gaussian source (source size $R = 1.2$ fm). The bands reflect the variation of the cutoff. The results obtained with quantum statistical effects and the Coulomb interaction are also shown for comparison.}
\label{CF}
\end{center}
\end{figure}

Notably, most of the previously mentioned meson-exchange and quark models fail to meet the requirements of Lorentz covariance. Meanwhile, the aforementioned chiral YN and YY interactions are all constructed based on the HB ChEFT. According to naive dimensional analysis, this non-relativistic expansion of the baryon propagator and Dirac spinors results in slow convergence. However, microscopic theories established within a relativistic framework have already demonstrated unique advantages in atomic and molecular systems, nuclear many-body systems, and single-baryon systems, such as improved theoretical self-consistency, faster chiral convergence, and better renormalization group invariance~\cite{Lu:2025syk}. Therefore, since 2016, the Beihang group has initiated the construction of YN and YY interactions based on the relativistic ChEFT~\cite{Li:2016mln}. So far, YN/YY interactions for systems with strangeness $S = -1$ to $-4$ have been constructed at leading order (LO). In the $S = -1$ sector, the low-energy constants were determined by fitting the low-energy scattering data, and a satisfactory description of these scattering data is obtained~\cite{Liu:2020uxi}, comparable to the NLO HB results. In addition, it was found that although the LO relativistic YN interaction is only constrained by low-energy data, it can reasonably well describe the latest J-PARC E40 data in the high-energy region~\cite {Song:2021yab}. In the $S = -2$ sector, one determined YN and YY interactions using lattice QCD simulation results~\cite{Liu:2022nec}, predicting an attractive $\Sigma\Sigma(I=2)$ interaction, which is not enough to generate a bound state. The computed femtoscopic correlation functions are in agreement with the ALICE measurements. In the $S = -3$ and $-4$ sectors, the lattice QCD phase shifts were described very well by the relativistic ChEFT~\cite{Liu:2020uxi}, and revealed the SU(3) flavor symmetry breaking among baryon-baryon interactions. Since $NN(I=1)$, $\Sigma\Sigma(I=2)$, and $\Xi\Xi(I=1)$ all belong to the same SU(3) irreducible representation, a quantitative examination of SU(3) flavor symmetry and its breaking can be performed by studying the femtoscopic correlation functions of $pp$, $\Sigma^+\Sigma^+$, and $\Xi^-\Xi^-$, as shown in Fig.~\ref{CF}.

Recently, the in-medium $\Lambda N$ interaction was studied within the relativistic Brueckner-Hartree-Fock framework, employing the above-obtained NN and YN interactions~\cite{Zheng:2025sol}. It was demonstrated that a consistent description of both the experimental cross-section data and the ‘empirical value' of the $\Lambda$ single-particle potential can be achieved. This contrasts with most non-relativistic studies, where higher-order two-body chiral forces are typically required.

\subsection{Theoretical progress on hypernuclear studies}
For decades,  many theoretical works have been
devoted to extracting information on the $\Lambda N$ interaction
from the structure of $\Lambda$ hypernuclei.
For these studies, one succeeded in obtaining information on the
two-body $\Lambda N$ interaction, including the spin-spin and spin-orbit parts.
One remaining important issue is to extract information on the
$\Lambda N -\Sigma N$ interaction which is related to
the charge symmetry breaking effect in the $\Lambda$ hypernuclei
and effective three-body $\Lambda NN$ force.
For this study, it is important to study neutron-rich 
$\Lambda$ hypernuclei. Because the total isospin of the
core nuclei is large,  the $\Lambda N-\Sigma N$coupling can
manifest itself in corresponding $\Lambda$ hypernuclei.
For instance, $^6$He, $^7$He, and $^8$He are neutron-rich nuclei. When one $\Lambda$ is added into these nuclei, due to the attractive $\Lambda N$  interaction,  bound states with respect to the lowest threshold have been found.
Then, it is possible to extract information on the
$\Lambda N -\Sigma N$ coupling effect from these structure.
Especially, in Ref.~\cite{Hiyama:2009ki},
it was emphasized that one could
extract information on charge symmetry-breaking effects from
the energies of $A=7$ iso-triplet $\Lambda$ hypernuclei.
Furthermore, recently, the J-PARC facility plans to
produce $^9_\Lambda$He, which is one neutron-rich $\Lambda$
hypernuclei. Therefore, energy spectra need to be predicted before measurement. 
In Ref.~\cite{Myo2023}, using the cluster orbital shell model,
 the energy spectra of these $\Lambda$ hypernuclei
as well as $A=6$ to 8 $\Lambda$ hypernuclei were studied.
Once new data on these $\Lambda$
hypernuclei are obtained, one could obtain information on the
$\Lambda N-\Sigma N$ coupling.
Then, one can go to the next step, to study the $S=-2$ sector, that is, the
$\Lambda \Lambda$ and $\Xi N$ interactions.
In the $S=-2$ sector, performing $YY$ scattering experiments isn't easy.
Therefore, it is crucial to study the structure of 
double $\Lambda$ hypernuclei and $\Xi$ hypernuclei.

\begin{figure}
\begin{center}
\includegraphics[scale=0.30]{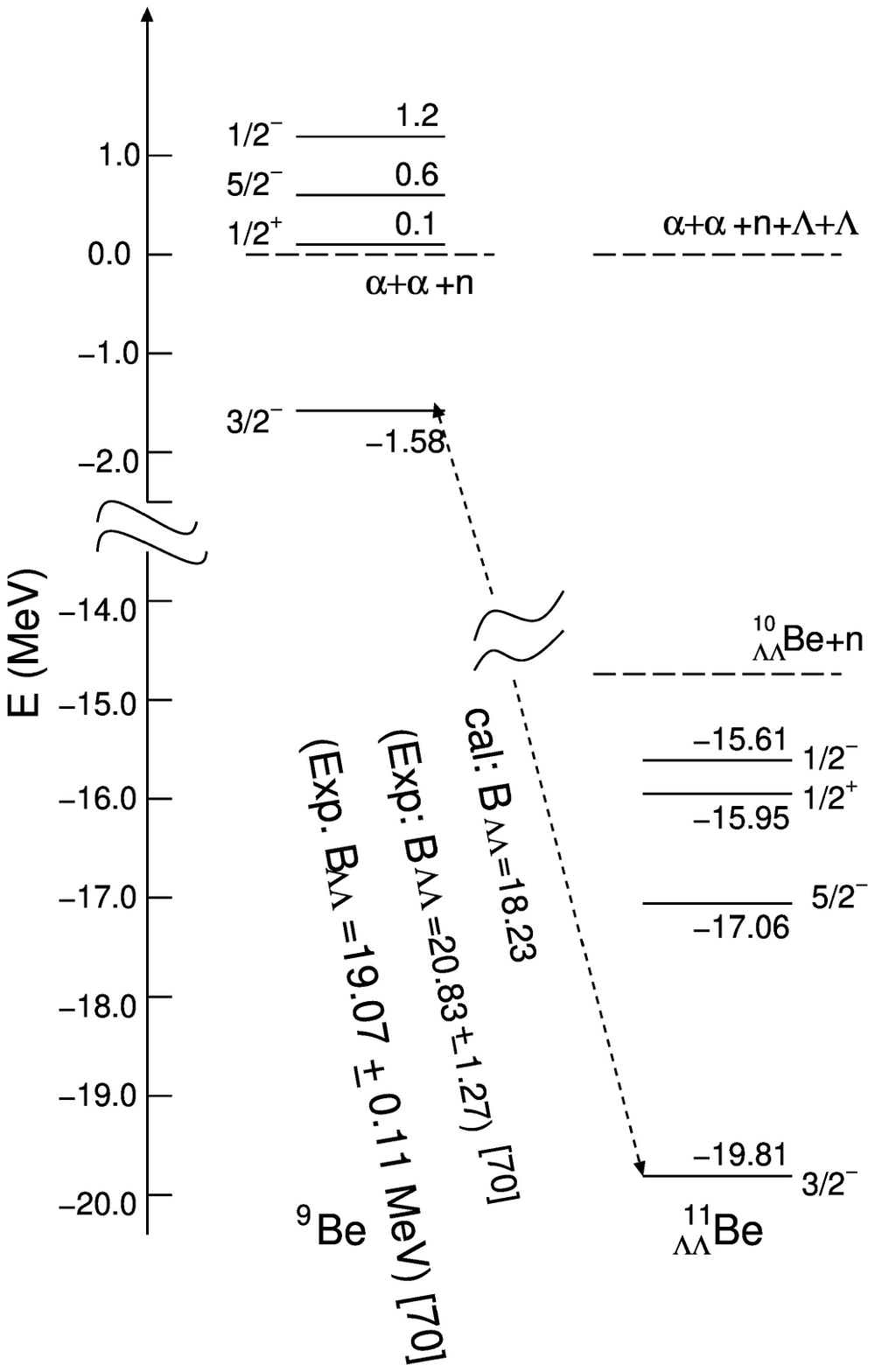}
\caption{Calculated energy spectra of the low-lying states of $^{11}_{\Lambda \Lambda}$Be together with those of the core
nucleus, $^9$Be}
\label{level-be11}
\end{center}
\end{figure}

For double $\Lambda$ hypernuclei, there were several data points by KEK-E176-E373:
the $2\Lambda$ separation energy ($B_{\Lambda \Lambda}$ of $^6_{\Lambda \Lambda}$He was $6.91 \pm 0.16$ MeV \cite{Hiyama2018} and
the observed $B_{\Lambda \Lambda}$ of
the excited state of $^{10}_{\Lambda \Lambda}$Be was $11.90 \pm 0.13$ MeV~\cite{Hiyama2018}. The theoretical aspects
of the $~^4_{\Lambda\Lambda}$H, $~^5_{\Lambda\Lambda}$He, and $~^6_{\Lambda\Lambda}$He had been studied in the Jacobi no-core shell mode (NCSM) with chiral interactions~\cite{Le:2021wwz}.
In addition, in 2009, $^{11}_{\Lambda \Lambda}$Be was observed and the
measured $B_{\Lambda \Lambda}$ was $20.83 \pm 1.27$ MeV. 
Afterwards, the search for the double $\Lambda$ hypernuclei by J-PARC E07 was performed at J-PARC. 
By this experiment, $B_{\Lambda \Lambda}$
of $^{11}_{\Lambda \Lambda}$Be was observed and 
the measured $B_{\Lambda \Lambda}$ was $19.07 \pm 0.11$ MeV. 
It should be noted that it is difficult to
identify states for the observed $B_{\Lambda \Lambda}$s
experimentally. Therefore, it is highly desirable to
interpret the states theoretically.
One successful example is  $^{11}_{\Lambda}$Be.
In 2009, new data of
double $\Lambda$ was observed as the Hida event \cite{Hiyama2018,Gal:2011zr}
as mentioned above
and there are two possibilities for interpretation:
One is $^{11}_{\Lambda \Lambda}$Be with $B_{\Lambda \Lambda}=20.83 \pm 1.27$ MeV, and the other is 
$^{12}_{\Lambda \Lambda}$Be with $B_{\Lambda \Lambda}=
22.43 \pm 1.21$ MeV. Furthermore, it is uncertain
whether this is an observation of a ground state or
an excited state.
To interpret this event, one of the present authors (E.H.) performed a study of the $\alpha +\alpha +n +\Lambda +\Lambda$
five-body cluster model by assuming that this event
was an observation of $^{11}_{\Lambda \Lambda}$Be
\cite{Hiyama:2010zzd,Hiyama2018}.
In this calculation, it is essential to reproduce 
the observed properties of 
all subcluster systems composed of two, three,
or four constituent particles.
In this calculation, the interactions are adjusted to reproduce energies of all subcluster systems, such as
$\alpha \alpha$, $\alpha \Lambda$, $\alpha n$, $\alpha \alpha \Lambda$,
$\alpha \alpha n$, $\alpha \Lambda \Lambda$ systems.
The energies of the ground state and the $2^+$ excited state
of $^{10}_{\Lambda \Lambda}$Be are
 in good agreement with the observed data with no adjustable parameter.
The calculated energy level is shown in Fig.~\ref{level-be11}.

The results of $^{11}_{\Lambda \Lambda}$Be are shown
in the energies of three bound excited states of
$1/2^-$, $1/2^+$ and $5/2^-$ as well as the ground state of $3/2^-$ below the lowest threshold of $^{10}_{\Lambda \Lambda}$Be.
Especially, it is interesting to see that the order of
the $1/2^+$ and $5/2^-$ state is reversed from the
core nucleus $^9$Be to the double $\Lambda$ hypernucleus
$^{11}_{\Lambda \Lambda}$Be.
The calculated $B_{\Lambda \Lambda}$ is 18.23 MeV
for the ground state of $^{11}_{\Lambda \Lambda}$Be,
which is less bound than the observed value of
20.83 MeV~\cite{Hiyama2018}.
The observed energy of $^{11}_{\Lambda \Lambda}$Be by the J-PARC E07 experiment is
$19.07 \pm 0.11$ MeV which does not contradict the result of Ref.~\cite{Hiyama2018}.
The analysis of the E07 experiment is still in progress.
It is expected to have more data on the $S=-2$ system in the
future.

It is also important to search for a bound $\Xi$
hypernucleus for the study of the $\Xi N$ interaction.
Experimentally, a bound $^{15}_{\Xi}$C($\Xi +^{14}$N)
hypernucleus was observed by emulsion~\cite{Nakazawa2015,Hiyama2018,Hayakawa2021,Yoshimoto2021}.
By this experimental data, one found that the $\Xi N$
interaction should be attractive.
In addition, the latest femtoscopic data from
$pA$ collisions at LHC~\cite{LHC} indicate that the spin-isospin averaged $\Xi N$ interaction is attractive at low energies.

Motivated by this observation, it is requested to
predict what kind of $\Xi$ hypernuclei should be observed as bound states experimentally and which $\Xi N$ spin-isospin term contributes to a bound state.

For this purpose, Ref.~\cite{Hiyama2008}
pointed out that the $^7_{\Xi}$H($\Xi+^6$He) and
$^{10}_{\Xi}$Li ($\Xi+^9$Be) hypernuclei should have
bound states to extract information on
the spin-isospin independent term of the $\Xi N$ interaction.
These two $\Xi$ hypernuclei are possible to be produced
by the $(K^-,K^+)$ reaction using $^7$Li and $^{10}$B targets.
Moreover, Ref.~\cite{Hiyama2020} performed
$NN\Xi$ and $NNN\Xi$ three- and four-body calculations 
using the Nijmegen potential such as ESC08c~\cite{ESC08c}
and the HAL-QCD potential based on first-principle lattice QCD simulations~\cite{HALQCD:2019wsz}.
In these two potentials, the partial wave contributions of the $\Xi N$ 
interaction, denoted by $^{2T+1,2S+1}S_J$ where
$T,S$, and $J$ stand for the total isospin, spin,
and angular momentum, are quite different from each other.
For ESC08c, the $^{33}S_1$ channel is strongly attractive to
have a bound state in the $\Xi N$ system, while $^{11}S_0$
and $^{31}S_0$ channels are weakly repulsive.
For the HAL QCD potential, $^{11}S_0$ is significantly attractive. On the other hand, the others are weakly repulsive or attractive.

The calculated energies of $NN\Xi$ and
$NNN\Xi$ are listed in Table~\ref{tab:nnx}.
The $NN\Xi$ system using the ESC08c potential
is bound for $(T,J^{\pi})=(1/2,3/2^+)$ with
respect to the $d+\Xi$ threshold with a binding energy of  7.2 MeV
Because this state receives contributions from the $N \Xi$ spin-1 channel,
which is strongly attractive in ESC08c.
On the other hand, the use of the HAL QCD potential does not
create any bound state in the $NN\Xi$ system.

For the $NNN\Xi$ system, we have two total isospins,
$T=0$ and 1.
For the $T=1$ system, the use of the ESC08c potential leads to
bound states for $0^+$ and $1^+$, while there is no bound state
using the HAL QCD potential.
For $T=0$, both the ESC08c and HAL QCD potentials
lead to a bound $J=0^+$ state, although 
the energies are dependent on the $\Xi N$ interaction employed. Meanwhile, three shallow bound states for the $NNN\Xi$ system with $(T,J^\pi)=(0,1^+)$, $(1,0^+)$ and $(1,1^+)$ are predicted in the Jacobi NCSM with the NLO(500) chiral interactions~\cite{Le:2021gxa}. It should be noted that some parameters of the chiral interactions used in Ref.~\cite{Le:2021gxa} have been retuned to the results of the original chiral baryon-baryon interaction, which was necessary because in the application to study $\Xi$ hypernuclei the $\Lambda\Lambda$ channel of the original potential had to be omitted.

By this calculation, one found that the $T=0$ $NNN\Xi$ system should be bound, which is likely to be produced in a future heavy-ion collision experiment. 

In the $S=-2$ sector, one has started to obtain information
on the $\Lambda \Lambda$  and $\Xi N$ interactions.
It is essential to obtain further information in this sector,
for instance, the $p$-wave  $\Lambda \Lambda$ interaction,
the spin-isospin terms of the $\Xi N$ interaction, etc.
New experiments on $S = -2$ hypernuclei, scattering experiments, and correlation functions together with improved theoretical calculations will help to shed further light on the $S = -2$ interactions.

\begin{table*} [tbh] 
\begin{center} 
\caption{Calculated binding energies (in units of MeV)
 of  $NN\Xi$ and $NNN\Xi$ with the
ESC08c and HAL QCD potentials with respect to the $d+\Xi$ and 
$^3{\rm H}/^3{\rm He}+\Xi$ thresholds, respectively.
}
\begin{tabular}{c|cc|cccc}
\hline\hline
                   &  $NN\Xi$     &                       &                                      &  $NNN\Xi$          
                      &                 &  \\
\hline 
$(T,J^{\pi})$    &  $(\frac{1}{2}, \frac{1}{2}^+)$ &  $(\frac{1}{2}, \frac{3}{2}^+)$   &  $(0, 0^+$)    & $(0, 1^+$)       & $(1, 0^+$)  &  $(1, 1^+)$   \\
\hline
ESC08c               & unbound                   &  $7.20$     & unbound                &  $10.20$               &    $3.55$                & $10.11$    \\
HAL QCD            & unbound                   &  unbound           & unbound               &  $0.36 (16)(26)$      &   unbound                     & unbound           \\
\hline\hline 
\end{tabular}
\label{tab:nnx}
\end{center}
\end{table*}

\section{Summary and outlook}\label{sec:sum}
Recent advances in hypernuclear physics have provided profound insights into the strangeness in nuclear systems, driven by experimental techniques and theoretical innovations. The measured properties of hypernuclei have been collected and evaluated in the online database (https://hypernuclei.kph.uni-mainz.de/). Light hypernuclei, such as the hypertriton ($^3_\Lambda$H), are pivotal benchmarks for probing the $\Lambda$-nucleon interaction. Notably, discrepancies in the binding energy of hypertriton between STAR and ALICE underscore the need for systematic studies to resolve these uncertainties. Future experiments at FAIR, J-PARC, and upgraded LHC facilities aim to deliver high-statistics data to clarify these inconsistencies and explore neutron-rich hypernuclei.

Charge symmetry breaking in the $A=4$ hypernuclear system has been an unsolved puzzle on both the experimental and theoretical sides for a considerable time. The most recent experimental measurements from the STAR Collaboration of $^4_{\Lambda}$H and $^4_{\Lambda}$He in Au+Au collisions at $\sqrt{s_{\rm NN}}$=3~GeV showed that the CSB effect in their excited states have a similar magnitude but opposite sign as their ground states. However, this measurement suffered from large uncertainties and could not draw any solid conclusions about the CSB puzzle. The STAR Collaboration has taken significantly larger statistics at $\sqrt{s_{\rm NN}}$ = 3 GeV. The analysis is ongoing, as one should expect to suppress the uncertainties in these new data. Measurements with electron and kaon beams are also in progress at MAMI-A1~\cite{A1:2025mjf}, JLab, and J-PARC. As demonstrated by the measurements of the STAR Collaboration, heavy-ion collisions at several GeV are excellent platforms for generating large statistics of light hypernuclei. Future experiments with heavy-ion beams, such as the High-Intensity Heavy-Ion Accelerator Facility (HIAF) \cite{Zhou:2022pxl} and the Facility for Antiproton and Ion Research (FAIR), are also possible for studying charge symmetry breaking in hypernuclei.

In the theoretical study of hyperon-nucleon and hyperon interactions, non-relativistic chiral effective potentials have been constructed up to the next-to-next-to-leading order. In contrast, the relativistic chiral potentials are only constructed at the leading order. Going to higher orders in both the non-relativistic and the relativistic frameworks requires more precise data to fix all the relevant low-energy constants. The recent measurement of differential cross sections by the J-PARC E40 experiment~\cite{J-PARCE40:2021bgw} and the femtoscopic correlation functions~\cite{ALICE:2021njx, ALICE:2019hdt, ALICE:2020mfd} are of tremendous help in this perspective. In addition to moving to higher orders in the two-body sector, studies of three-body hyperon-nucleon-nucleon interactions can play a crucial role in understanding various relevant issues, such as the hyperon puzzle in neutron stars. 

Bare interactions provide necessary inputs in studies of hypernuclei. Conventional theoretical methods, such as the GEM method, need to be extended to treat a wider range of hadrons. Currently, it can treat five body systems. Another promising approach is the lattice effective field theory approach~\cite{Hildenbrand:2024ypw}. Relativistic few-body calculations utilizing relativistic chiral forces are in progress and may provide unique insights into many unsolved problems. For heavy nuclei/hypernuclei out of reach of ab initio methods, one can first derive effective potentials using bare chiral hyperon/nucleon forces, and then use them as inputs with density functional theories~\cite{Yang:2021akb}.

\section*{Acknowledgements}
J.H.C. is supported by the National Key R\&D Program of China under Grant No. 2022YFA1604900 and by the National Natural Science Foundation of China under Grant No. 12025501.
L.S.G is partly supported by the National Key R\&D Program of China under Grant No. 2023YFA1606703 and the Natural Science Foundation of China under Grant No. 12435007. Z.W.L. acknowledges support from the National Natural Science Foundation of China under Grant No.12405133, No.12347180, China Postdoctoral Science Foundation under Grant No.2023M740189, and the Postdoctoral Fellowship Program of CPSF under Grant No.GZC20233381.

\bibliography{citations,lit_phys_hyp1,lit_phys_hyp2}

\end{document}